\pdfoutput=1
\documentclass[PAPER, atlasdraft=false, cernpreprint, USenglish, texmf, orcidlogo]{atlasdoc}
\usepackage[subcaption]{atlaspackage}
\usepackage{comment}
\usepackage{atlasbiblatex}

\usepackage{atlasphysics}
\graphicspath{{logos/}{figures/}}

\usepackage{ANA-SUSY-2020-08-PAPER-defs}

\AtlasTitle{Search for long-lived particles using displaced vertices of oppositely charged leptons in $140~\mathrm{fb^{-1}}$ of $\mathit{pp}$ collisions at $\sqrt{s} = 13$~TeV with the ATLAS detector}

\AtlasAbstract{%
A search is presented for long-lived particles decaying into an oppositely charged lepton pair, $\mu^{+}\mu^{-}$, $e^{+}e^{-}$, or $e^{\pm}\mu^{\mp}$, that form a vertex within the inner tracking system of the ATLAS detector at the Large Hadron Collider, displaced from the primary proton--proton interaction region.
The analysis uses the 140~$\mathrm{fb^{-1}}$ of Run-2 data collected at $\sqrt{s}=13$~TeV by the ATLAS experiment in 2015--2018.
The results of the analysis are interpreted in the context of three benchmark models covering masses from 0.1~to~2.2~TeV and a range of mean proper lifetimes times the speed of light from 1~to~10\,000~mm.
The first model is a generic $Z'$ boson pair-produced by a new heavy scalar, with the $Z'$ decaying into lepton pairs.
The remaining two models are $R$-parity violating supersymmetric models in which the lightest neutralino $\tilde{\chi}^{0}_{1}$ decays into $\ell^{+}\ell^{'-}\nu$ ($\ell, \ell^{'} = e$, $\mu$).
The models differ by the mode of production of the $\tilde{\chi}^{0}_{1}$, which can be produced via the decay of pairs of gluinos or of pairs of charginos and neutralinos ($\tilde{\chi}_{1}^{\pm}\tilde{\chi}_{1}^{0}$, $\tilde{\chi}_{1}^{\pm}\tilde{\chi}_{2}^{0}$, or $\tilde{\chi}_{2}^{0}\tilde{\chi}_{1}^{0}$).
Although each benchmark sample includes pair-produced LLPs, only a single vertex is required to be reconstructed.
No dilepton displaced vertex candidate is observed and the results are presented as upper limits on the production cross-sections.
This analysis sets leading limits on the production cross-sections for multiple models, including parameter space that has never been directly probed.
}

\AtlasRefCode{SUSY-2020-08}

\PreprintIdNumber{CERN-EP-2025-293}

\arXivId{2601.05664}

\HepDataRecord{169597}

\AtlasJournal{Phys.\ Lett.\ B.}
\AtlasJournalRef{\PLB 878 (2026) 140497}
\AtlasDOI{10.1016/j.physletb.2026.140497}

\hypersetup{pdftitle={ATLAS document},pdfauthor={The ATLAS Collaboration}}

\begin{document}

\maketitle

%

%
%

%
%

\section{Introduction}
\label{sec:intro}

Searches for long-lived particles (LLPs) remain a well-motivated endeavor in the collider physics community.
LLPs are a natural occurrence in many extensions of the Standard Model (SM), including Hidden Valley~\cite{Han:2007ae} models and various supersymmetric (SUSY) models like split supersymmetry, $R$-parity violating (RPV) SUSY~\cite{Dreiner:1997uz,Barbier:2004ez}, and gauge-mediated SUSY breaking models~\cite{Nappi:1982hm,AlvarezGaume:1981wy,Dine:1981gu}.
Their ubiquity is in part a result of the multiple possible means by which long lifetimes may arise, such as small couplings, small mass-splittings, and approximate symmetries.
A neutral LLP can be detected indirectly by reconstructing its decay products, which may form a common vertex that is macroscopically displaced from the primary collision vertex.

This Letter presents the search for a heavy, long-lived, neutral particle whose decays include a dilepton pair ($\mu^{+}\mu^{-}$, $e^{+}e^{-}$, or $e^{\pm}\mu^{\mp}$) producing a displaced vertex (DV) that is reconstructed within the inner tracking system of the ATLAS detector~\cite{PERF-2007-01} at the Large Hadron Collider (LHC)~\cite{Evans:2008zzb}.
In this Letter, ``leptons'' refers to electrons and muons only.
LLP decays into taus are not modeled or targeted by this analysis.
The analysis uses 140~\ensuremath{\text{fb}^{-1}}\xspace of Run-2 data collected in proton--proton ($pp$) collisions at $\sqrt{s}=13$~\TeV between 2015 and 2018.
Requirements are placed on reconstructed vertices to strongly suppress SM background, allowing for sensitivity to signal processes with small cross-sections.

Three benchmark models are considered.
The first model simulates the production of a neutral, heavy scalar particle $S$ that decays into a pair of long-lived $Z'$ bosons that in turn decay into charged leptons.
The second and third models are each simplified models motivated by the Minimal Supersymmetric Standard Model (MSSM) featuring a long-lived lightest neutralino that acquires its long lifetime due to a small RPV coupling and decays into two charged leptons and a neutrino.
One of these MSSM-inspired RPV models features direct electroweakino production ($\tilde{\chi}_{1}^{\pm}\tilde{\chi}_{1}^{0}$, $\tilde{\chi}_{1}^{\pm}\tilde{\chi}_{2}^{0}$, or $\tilde{\chi}_{2}^{0}\tilde{\chi}_{1}^{0}$) and the other features pair-production of gluinos ($\tilde{g}\tilde{g}$), which each decay into two quarks and the long-lived neutralino.

Requirements on collision events are not tailored to these specific models and no requirements are imposed on missing transverse momentum or jets.
To maintain as much model independence as possible the reconstructed vertices are permitted to have extra tracks in addition to the two signal lepton tracks.

A previous search by the ATLAS Collaboration for the same signature was performed on 32.8~\ensuremath{\text{fb}^{-1}}\xspace of data collected in 2015 and 2016, using a squark production model rather than the gluino or electroweakino production models used here~\cite{SUSY-2017-04}.
The ATLAS Collaboration also searched for dimuon vertices reconstructed using only muon-spectrometer tracks, allowing longer proper lifetimes to be targeted, based on the same subset of Run-2 data~\cite{EXOT-2017-03}.
Additionally, the ATLAS Collaboration searched for delayed dielectron and diphoton vertices reconstructed using only calorimeter information, targeting masses consistent with the $Z$ or Higgs bosons~\cite{SUSY-2020-28}.
The CMS Collaboration searched for displaced dilepton vertices in the inner tracking volume at $\sqrt{s}=8$~\TeV~\cite{CMS-EXO-12-037}.
Using data collected at $\sqrt{s}=13$~\TeV, the CMS Collaboration searched for pairs of collimated, displaced muons reconstructed using only information from the muon system in events with missing transverce momentum~\cite{CMS-EXO-20-010} and for lower mass LLPs decaying into muon pairs using dedicated trigger streams~\cite{CMS-EXO-20-014}.
Also using data collected at $\sqrt{s}=13$~\TeV, the LHCb Collaboration searched for lower mass dimuon vertices in 5.1~\ensuremath{\text{fb}^{-1}}\xspace of data~\cite{LHCbDiMuon}.
Finally, the CMS Collaboration released results for a Run-3 search targeting displaced dimuon vertices using 36.6~\ensuremath{\text{fb}^{-1}}\xspace of $\sqrt{s}=13.6$~\TeV data~\cite{CMS-EXO-23-014}.
This is the first published analysis from the ATLAS Collaboration to use the full Run-2 data at $\sqrt{s}=13$~\TeV for a search for dilepton displaced vertices and is the only search using full Run-2 data to include dimuon, dielectron, and electron-muon displaced vertices.


\section{ATLAS detector}
\label{sec:detector}

The ATLAS detector at the LHC covers nearly the entire solid angle around the collision point.\footnote{\AtlasCoordFootnote}
It consists of an inner tracking detector surrounded by a thin superconducting solenoid, electromagnetic and hadronic calorimeters,
and a muon spectrometer incorporating three large superconducting air-core toroidal magnets.

The inner-detector system (ID) is immersed in a \qty{2}{\tesla} axial magnetic field
and provides charged-particle tracking in the range \(|\eta| < 2.5\).
The high-granularity silicon pixel detector covers the vertex region and typically provides four measurements per track,
the first hit generally being in the insertable B-layer (IBL) installed before Run~2~\cite{ATLAS-TDR-19,PIX-2018-001}.
It is followed by the SemiConductor Tracker (SCT), which usually provides eight measurements per track.
The barrel pixel layers, which are positioned at radii of 33.3, 50.5, 88.5, and 122.5~mm are of particular relevance to this search.
The radial positions of the innermost and outermost SCT barrel layers are 299~mm and 514~mm, respectively.
These silicon detectors are complemented by the transition radiation tracker (TRT),
which enables radially extended track reconstruction up to \(|\eta| = 2.0\).
The TRT also provides electron identification information
based on the fraction of hits (typically 30 in total) above a higher energy-deposit threshold corresponding to transition radiation.

The calorimeter system covers the pseudorapidity range \(|\eta| < 4.9\).
Within the region \(|\eta|< 3.2\), electromagnetic calorimetry is provided by barrel and
endcap high-granularity lead/liquid-argon (LAr) calorimeters,
with an additional thin LAr presampler covering \(|\eta| < 1.8\)
to correct for energy loss in material upstream of the calorimeters.
Hadronic calorimetry is provided by the steel/scintillator-tile calorimeter,
segmented into three barrel structures within \(|\eta| < 1.7\), and two copper/LAr hadronic endcap calorimeters.
The solid angle coverage is completed with forward copper/LAr and tungsten/LAr calorimeter modules
optimized for electromagnetic and hadronic energy measurements, respectively.

The muon spectrometer (MS) comprises separate trigger and
high-precision tracking chambers measuring the deflection of muons in a magnetic field generated by the superconducting air-core toroidal magnets.
The field integral of the toroids ranges between \num{2.0} and \qty{6.0}{\tesla\metre}
across most of the detector.
Three layers of precision chambers, each consisting of layers of monitored drift tubes, cover the region \(|\eta| < 2.7\),
complemented by cathode-strip chambers in the forward region, where the background is highest.
The muon trigger system covers the range \(|\eta| < 2.4\) with resistive-plate chambers in the barrel, and thin-gap chambers in the endcap regions.

The luminosity is measured mainly by the LUCID--2~\cite{LUCID2} detector that records Cherenkov light produced in the quartz windows of photomultipliers located close to the beampipe.

Events were selected by the first-level trigger system implemented in custom hardware,
followed by selections made by algorithms implemented in software in the high-level trigger~\cite{TRIG-2016-01}.
The first-level trigger accepted events from the \qty{40}{\MHz} bunch crossings at a rate close to \qty{100}{\kHz},
which the high-level trigger further reduced in order to record complete events to disk at about \qty{1.25}{\kHz}.

A software suite~\cite{SOFT-2022-02} is used in data simulation, in the reconstruction
and analysis of real and simulated data, in detector operations, and in the trigger and data acquisition
systems of the experiment.


%
%

\section{Benchmark models}
\label{sec:mods}

This analysis is motivated by the distinct experimental signature and minimal background associated with the displaced dilepton vertices, rather than by a specific extension of the SM that could produce such a signature.
Three models that could produce a leptonically decaying LLP are used as benchmark models to evaluate the sensitivity of the search.
All three models contain pair-produced LLPs, but this analysis does not require multiple LLP decays to be reconstructed.
The analysis is therefore sensitive to models containing a single LLP.

\begin{figure}%
\centering
\begin{subfigure}{0.28\linewidth}
\includegraphics[width=\textwidth]{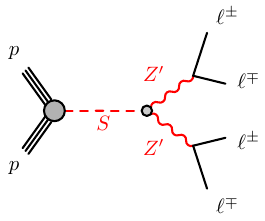}
\caption{}\label{fig:zprime}
\end{subfigure}
\hspace{0.5 cm}
\begin{subfigure}{0.28\linewidth}
\includegraphics[width=\linewidth]{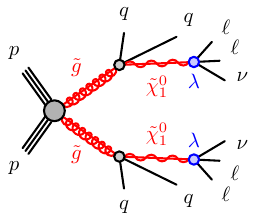}
\caption{}\label{fig:gg}
\end{subfigure}
\hspace{0.5 cm}
\begin{subfigure}{0.28\linewidth}
\includegraphics[width=\linewidth]{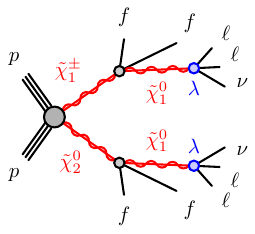}
\caption{}\label{fig:ew}
\end{subfigure}
\caption{
Diagrams for (a) a pair of $Z'$ produced via a scalar and for a pair of $\tilde{\chi}_{1}^{0}$, each produced via (b) a gluino or (c) an electroweakino.
For the benchmark models used, the $Z'$ and $\tilde{\chi}_{1}^{0}$ are the LLPs while the scalars, gluinos, and higher-mass electroweakinos ($\tilde{\chi}_{1}^{\pm}$ and $\tilde{\chi}_{2}^{0}$) decay promptly. The symbol $\lambda$ represents a RPV coupling as described in the text.
}
\label{fig:susyfeynmandiagram}
\end{figure}

The first benchmark model (``$Z'$ model'') is a simplified model containing a heavy scalar $S$ that decays into a pair of long-lived $Z'$.
A diagram of this process is shown in Figure~\ref{fig:zprime}.
Prompt decays of a $Z'$ to leptons are heavily constrained~\cite{ParticleDataGroup:2024cfk}, but long-lived $Z'$ remain viable.
In the simplified model used here the lifetime of the $Z'$ is treated as a free parameter and it is forced to decay into lepton pairs.

The second and third benchmark signal models are both RPV MSSM simplified models where the lightest neutralino $\tilde{\chi}^{0}_{1}$ is the lightest supersymmetric particle (LSP).
$R$-parity violation allows the LSP neutralino to decay, possibly with a sufficiently long lifetime to produce a displaced vertex~\cite{Barbier:2004ez}.
The two SUSY models implement different production mechanisms of the long-lived neutralino.
In both models, the $\tilde{\chi}_{1}^{0}$ decays via RPV couplings $\lambda_{121}$ or $\lambda_{122}$~\cite{Dreiner:1997uz} into two charged leptons and a neutrino.
The $\lambda_{121}$ coupling results in $ee\nu$ and $e\mu\nu$ decays each with equal branching ratios and the $\lambda_{122}$ coupling results in $e\mu\nu$ and $\mu\mu\nu$ decays each with equal branching ratios.
No requirement on jets, missing transverse momentum, or other leptons is imposed so that the search is as model-independent as possible.
The first MSSM model (``Gluino model'') implements the pair production of gluinos, where the gluino decays promptly via $\tilde{g} \rightarrow{} qq\tilde{\chi}_{1}^{0}$ (Figure~\ref{fig:gg}).
The second MSSM model (``EWkino model'') implements electroweakino pair production ($\tilde{\chi}_{1}^{\pm}\tilde{\chi}_{1}^{0}$, $\tilde{\chi}_{1}^{\pm}\tilde{\chi}_{2}^{0}$, or $\tilde{\chi}_{2}^{0}\tilde{\chi}_{1}^{0}$), an example of which is shown in Figure~\ref{fig:ew}.
The charginos and neutralinos are assumed to be pure higgsino-like and are nearly mass degenerate, with a 1~\GeV mass splitting between the lightest neutralino and both the chargino and heavier neutralino.
If a $\tilde{\chi}_{1}^{\pm}$ or $\tilde{\chi}_{2}^{0}$ is produced, it decays promptly via $\tilde{\chi} \rightarrow{} ff\tilde{\chi}_{1}^{0}$, where $f$ is any kinematically allowed fermion.

\section{Data and simulated events}
\label{sec:samps}

The data used in this analysis were collected from $\mathit{pp}$ collisions at $\sqrt{s}=13$~\TeV during Run 2 (2015--2018), corresponding to an integrated luminosity of 140.1$\pm$1.2~\ensuremath{\text{fb}^{-1}}~\cite{DAPR-2021-01}.
All backgrounds are estimated using collision data.
A selection of muon and electron events from the same data were used for calculating corrections to the Monte Carlo (MC) modeling of lepton efficiencies and their associated systematic uncertainties.

The matrix element (ME) calculations for the $Z'$ model are performed at next-to-leading-order (NLO) by \MADGRAPH[2.6.7]~\cite{Alwall:2014hca} interfaced with \PYTHIA[8.244]~\cite{Sjostrand:2014zea} using the A14 set of tuned parameters~\cite{ATL-PHYS-PUB-2014-021} with the \NNPDF[3.0nlo] set of parton distribution functions (PDF)~\cite{Ball:2014uwa}.
Signal MC samples are produced for scalar masses from 500~\GeV to 1.5~\TeV, several $Z'$ masses ranging from 100~to~700~\GeV depending on the scalar mass, and $Z'$ mean proper lifetimes times speed of light ($c\tau$) of 10,~30,~100,~300,~and~1000~mm.
The detector response to each simulated event for all simulated samples is modeled by GEANT4~\cite{Agostinelli:2002hh,SOFT-2022-02}.

The ME calculations for the Gluino model and the EWkino model samples are performed at leading-order (LO) by \MADGRAPH[2.8.1]~\cite{Alwall:2014hca} interfaced with \PYTHIA[8.244] using the A14 set of tuned parameters with the \NNPDF[3.0nlo] PDF set.
The MEs include up to two extra parton emissions.
Signal MC samples for the Gluino model are produced for $\tilde{g}$ masses ranging from 1.5~to~2.5~\TeV and several $\tilde{\chi}_{1}^{0}$ masses.
The heaviest $\tilde{\chi}_{1}^{0}$ mass used for each $\tilde{g}$ mass is 300~\GeV less than the $\tilde{g}$ mass and the lightest is always 100~\GeV.
Samples for the EWkino model are produced for $\tilde{\chi}_{1}^{0}$ masses ranging from 100~\GeV to 1.3~\TeV.
For both models, separate samples are produced for the same five $c\tau$ for the $\tilde{\chi}_{1}^{0}$ as were used for the $Z'$.

The effect of multiple interactions in the same and adjacent bunch crossings (pileup) was modeled by overlaying the simulated hard-scattering event with inelastic \(pp\) events generated with \PYTHIA[8.186]~\cite{Sjostrand:2007gs} using the \NNPDF[2.3lo] PDF set and the A3 set of tuned parameters~\cite{ATL-PHYS-PUB-2016-017}.
Events in the signal MC samples are weighted such that the pileup distribution in MC closely resembles that in data.

A $t\bar{t}$ sample is used for validation of the procedure for estimating the background from DV formed by random track crossing (Section~\ref{sec:bkg}) and as a source of $K_{S}$ used for a systematic study of combined track reconstruction and vertex reconstruction efficiency for LLPs (Section~\ref{sec:systs}).
This sample is simulated using \POWPY[8.230]~\cite{Sjostrand:2014zea, Frixione:2007vw} using the A14 set of tuned parameters with the \NNPDF[3.0nlo] PDF set.

Two $Z$ boson MC samples ($Z\rightarrow e^{+}e^{-}$ and $\mu^{+}\mu^{-}$) are also used to compare to data when deriving multiplicative factors to correct for imperfect modeling of electron and muon detection efficiency.
These samples are simulated using \POWPY[8.186]~\cite{Alioli:2008gx} using the \AZNLO~\cite{STDM-2012-23} set of tuned parameters with the CT10 PDF set~\cite{Lai:2010vv}.


\section{Object reconstruction}
\label{sec:obreco}

The standard track (ST) reconstruction algorithm reconstructs the trajectories of charged particles (tracks) from hits in the ID\@.
Requirements placed on the transverse ($d_{0}$) and longitudinal ($z_{0}$) impact parameters limit the reconstruction efficiency of tracks from displaced decays.
To mitigate this reduction in efficiency, tracks from a large radius track (LRT) reconstruction algorithm~\cite{ATL-PHYS-PUB-2017-014} are also used.
The LRT algorithm reconstructs tracks using hits that are not associated with any track from ST reconstruction.
Relevant differences between the track requirements for the ST and LRT algorithms are compared in Table~\ref{table:lrt_comparison}.
Of particular note is the relaxation of the requirements for LRT reconstruction on $d_{0}$ and $z_{0}$, as well as the maximum number of shared silicon modules and the minimum number of unshared silicon hits between two tracks.
These loosened requirements increase the reconstruction efficiency for tracks not originating from the primary vertex (PV).

The PV reconstruction algorithm~\cite{ATL-PHYS-PUB-2015-026} reconstructs vertices using tracks that have parameters consistent with originating from the beam collision region in the $x$-$y$ plane and is therefore not optimized for displaced decays.
A dedicated DV reconstruction algorithm is used to reconstruct vertices at large distances from the PV\@.
The algorithm~\cite{ATL-PHYS-PUB-2019-013} uses tracks to reconstruct DVs within the pixel and SCT subsystems of the ID\@.
All tracks from both the ST and LRT reconstruction algorithms that have $\pT > 1$~\GeV, $|d_0| < 300$~mm, and $|z_0| < 1500$~mm are considered for DV reconstruction.
No calorimeter or MS information associated with the track is used, and no corrections are made to the tracks based on their lepton association.
The reconstruction algorithm begins by forming a prospective vertex from each possible combination of two tracks, using only tracks not included in collision vertices.
To reject tracks not originating from a DV, tracks are only included in the DV if they contain no pixel or SCT hits between the DV and the PV and if they have a hit in the first pixel or SCT layer as the track emerges from the DV.\footnote{A dead pixel or disabled module is treated as a ``hit'' in both cases, i.e. a dead pixel in between the vertex and PV would cause the vertex to be rejected but a dead pixel in the first layer as the track emerges from the DV would satisfy the requirement for a hit.\label{fn:VSIreq}}
Additionally, at least one of the two tracks in the prospective vertex must satisfy $|d_0| > 2$~mm.
Vertices with higher track multiplicity are formed by iteratively merging nearby vertices if this improves the goodness of fit of all tracks to the common vertex.
Tracks may then be pruned from the merged vertex to further improve the goodness of fit.

\begin{DIFnomarkup}
\begin{table}[tbp]
\centering
\caption{Comparison of requirements between the standard track and large radius track reconstruction algorithms~\cite{ATL-PHYS-PUB-2017-014}.}
\label{table:lrt_comparison}
\begin{tabular}{ l  r  r }
\hline
\hline
& ST & LRT \\ [0.5ex]
\hline
Maximum $|d_{0}|$ [mm] & 10 & 300 \\
Maximum $|z_{0}|$ [mm] & 250 & 1500 \\
Maximum $|\eta|$ & 2.7 & 5 \\
Maximum shared silicon modules & 1 & 2 \\
Minimum unshared silicon hits& 6 & 5 \\
Minimum silicon hits & 7 & 7\\
\hline
\hline
\end{tabular}
\end{table}
\end{DIFnomarkup}

Leptons are reconstructed with the standard ATLAS reconstruction algorithms for electrons~\cite{EGAM-2018-01,EGAM-2021-01} and muons~\cite{MUON-2018-03}.
Calorimeter information is combined with track information from the ST and LRT reconstruction algorithms to reconstruct and identify electrons.
In the case when several tracks match the same calorimeter energy deposit, the track with the best momentum match to the calorimeter energy is chosen to reconstruct the electron.
An additional algorithm~\cite{EGAM-2018-01} refines the parameters of tracks corresponding to electrons, accounting for energy loss due to bremsstrahlung.
The refined tracks are not used in the DV reconstruction algorithm or to determine the kinematic properties of the DV, so the updated information is only used in this analysis for identifying the triggering leptons.

Muons must be of the combined type defined in Ref.~\cite{MUON-2018-03}.
Combined muons are reconstructed by pairing tracks reconstructed using only MS information with ID tracks (either reconstructed by the ST or LRT algorithms).
A combined fit is then performed using both ID and MS hits and accounting for energy loss in the calorimeters.
The combined fitting procedure allows for a reassignment of MS hits based on the combined fit trajectory.

\section{Event selection}
\label{sec:evsel}

Events with dilepton DVs are selected via a series of criteria, either applied to the event, the constituent leptons, or the DVs themselves.
The event selection begins by selecting collision events with leptons that may be displaced and requiring the presence of a DV\@.
Criteria are then applied to the displaced leptons and their associated DVs to isolate potential signal events.

\subsection{Event level selection criteria}
\label{sec:presel}

Events are required to have a PV with at least two associated tracks.
The PV is selected as the $\mathit{pp}$ collision vertex with the largest \(\Sigma \pT^2\), the squared transverse momentum of all associated tracks with $\pT > \qty{0.5}{\GeV}$.
Additionally, events must be recorded during time periods where detector subsystems were functioning under good operating conditions~\cite{DAPR-2018-01}.

Events are selected using triggers that do not rely on ID track information.
Track reconstruction algorithms used by the Run-2 triggers do not include an equivalent of the LRT algorithm; therefore, a requirement for the presence of an ID track can suppress trigger efficiency for displaced leptons.
The ``MS only'' muon trigger selects an MS track with $\pt > 60$~\GeV and $|\eta| < 1.05$~\cite{TRIG-2018-01}.
For electrons two triggers are used: a single photon trigger and a diphoton trigger~\cite{TRIG-2018-05}.
The use of a photon trigger is appropriate to target displaced electrons because electrons and photons leave similar energy deposits in the electromagnetic calorimeter.
The single photon trigger requires the reconstructed object satisfy $\pt > 140$~\GeV, while the diphoton trigger imposes a softer requirement, $\pt > 50$~\GeV, for both objects.
No requirement is placed on $\eta$ for either photon trigger other than the implicit bounds set by the geometry of the electromagnetic calorimeter.
The combined trigger efficiency is typically above 80\% for all vertex types for decays within the fiducial region.
Individually, the electron triggers generally have a higher efficiency than the muon trigger due to their larger pseudorapidity coverage.
Events selected by the trigger are required to include a reconstructed lepton which spatially matches the location of the object selected by the trigger.
The triggering lepton must also satisfy both $|d_0| >$ 2~mm and a requirement on its $\pT$ that matches the trigger requirement.
If the event was selected by the diphoton trigger, two leptons must satisfy the trigger requirements.

A cosmic veto is applied to suppress cosmic-ray muon background contributions.
A cosmic-ray muon may be mis-reconstructed as a pair of back-to-back and oppositely charged muons.
Calorimeter deposits resulting from muon bremsstrahlung may also cause a muon to be reconstructed as an electron.
Back-to-back tracks can be distinguished by values of $\rcos$ that are close to zero, where $\Sigma\eta=\eta_{1}+\eta_{2}$ and $\Delta\phi=\phi_{1}-\phi_{2}$ for the $\eta$ and $\phi$ of the two tracks.
Imposing the requirement $\rco>0.01$ on all $\mu\mu$ pairs and $\mu{}e$ pairs in the event effectively eliminates these cosmic-ray muon events.
The number of signal events eliminated by the cosmic veto does not exceed 1\%.

\subsection{Lepton selection criteria}
\label{sec:lepsel}

\begin{figure}%
\centering
\includegraphics[width=0.9\linewidth]{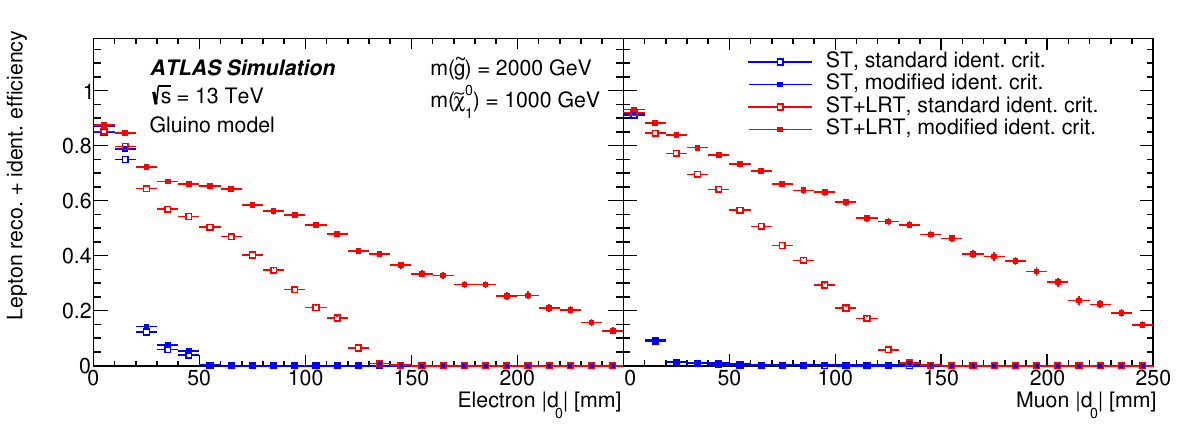}
\caption{The efficiency for reconstructing and correctly identifying a simulated lepton from an LLP decay as a function of $d_{0}$ for (left) electrons and (right) muons for ST and ST+LRT, with standard and modified working points. The efficiencies were determined using gluino-produced $\tilde{\chi}_{1}^{0}$ samples with a $\tilde{g}$ mass of 2000~\GeV and a $\tilde{\chi}_{1}^{0}$ mass of~1000 \GeV, averaged over the five $c\tau$ from 10 to 1000~mm. }
\label{fig:modifiedID}
\end{figure}

To qualify as signal electrons, reconstructed electrons must satisfy the VeryLooseLLH identification requirements defined in Ref.~\cite{EGAM-2018-01} but with requirements on $d_0$ and on the number of pixel hits associated with the track removed.
Signal muons must satisfy the Loose identification requirements defined in Ref.~\cite{MUON-2018-03} but with the requirements on the number of pixel hits associated to the track removed.
Figure~\ref{fig:modifiedID} shows the efficiency for an electron or a muon from an LLP decay in a benchmark sample to be both reconstructed and satisfy lepton identification criteria.
The efficiencies are shown using only the ST reconstruction algorithm versus both ST and LRT, and the standard versus modified lepton identification criteria.
Including the LRT and modified identification criteria greatly improves the efficiency.
Multiplicative factors to correct for the mismodeling of lepton identification and triggering efficiencies in signal samples are also applied, as discussed in Section~\ref{sec:systs}.
An overlap removal to remove ambiguity between electrons, muons, and jets is applied similar to that described in Ref.~\cite{HIGG-2019-09}, with the exception that the $\Delta R$ requirement between electrons and muons is not applied to allow for more highly collimated signal leptons.

Some analysis requirements are applied to the lepton's kinematic parameters, while others are applied to the kinematic parameters of the ID track, which may be different.
The use of ID track parameters, which are those available to the DV reconstruction algorithm, simplifies the calculation of systematic uncertainties.
The kinematic parameters of the ID track are used for the baseline track requirements $\pt>10~\GeV$ for both lepton flavors and $|\eta| < 2.47$~($2.5$) for electrons (muons).
The kinematic parameters of the leptons themselves are used for lepton identification, charge requirements, the cosmic veto, and trigger matching, the last of which is discussed in Section~\ref{sec:dvsel}.

The combined fit of the muon track based on both ID and MS hits must satisfy ${\chi^{2}/\textrm{dof}<10}$.
The displaced decay of a charged hadron to a muon may be incorrectly reconstructed as a combined muon if the ID track from the hadron is matched to the MS track of the muon by the reconstruction software.
In such cases, the ID and MS components are often poorly aligned, so the requirement on $\chi^{2}$ effectively removes most muons originating from charged hadron decays with negligible impact on signal muons in benchmark samples.

\subsection{Displaced vertex selection criteria}
\label{sec:dvsel}
\begin{sloppypar}
To suppress background, signal DVs must have transverse displacement from the PV satisfying ${R_{xy} = \sqrt{(x_{\text{DV}} - x_\text{{PV}})^{2} + (y_{\text{DV}} - y_\text{{PV}})^{2}} > 2~\text{mm}}$.
The DV must also satisfy ${\sqrt{(x_{\text{DV}} )^{2} + (y_{\text{DV}})^{2}} < 300~\text{mm}}$ and ${|z_{\text{DV}}| < 300~\text{mm}}$ (both measured relative to the ATLAS detector origin), defining a fiducial region in the barrel of the ID bounded by the innermost SCT layer.
DVs are rejected if they are within disabled pixel modules~\cite{SUSY-2016-08}.
A separate ``material veto''~\cite{PERF-2015-07}, designed to suppress displaced vertices from photon conversions, is applied to DVs that contain an electron and are within the tracking layers or the support structures.
Following the strategy used in previous ATLAS DV analyses~\cite{SUSY-2018-33,SUSY-2018-13,EXOT-2021-32}, DVs are required to be reconstructed with $\chi^2 /\textrm{dof} < 5$.
An event is permitted to contain more than one signal DV.
\end{sloppypar}

At least two of the ID tracks in the DV must be associated with oppositely charged leptons that satisfy the modified identification criteria discussed in Section~\ref{sec:lepsel}.
One of the leptons in a DV must be the lepton selected by the trigger (both leptons in the case of the diphoton trigger) and satisfy the stricter kinematic requirements discussed in Section~\ref{sec:presel}.
Finally, the invariant mass of the ID tracks of the lepton pair forming the DV must satisfy $m_{\text{DV}} > 12$~\GeV to suppress backgrounds from SM particles.
To allow reinterpretation using other models, additional tracks (including leptonic tracks) are permitted in the DV despite not being expected in the benchmark models.
Enforcing a requirement of exactly two leptonic tracks in a DV would remove $<1\%$ of signal vertices in the benchmark models.
In the case of a DV containing three or more leptonic tracks, each pair of tracks is treated as a potential signal DV, with separate trigger matching and mass requirements for each track pair.

\begin{figure}[tbp]
\centering
\includegraphics[width=1.0\textwidth]{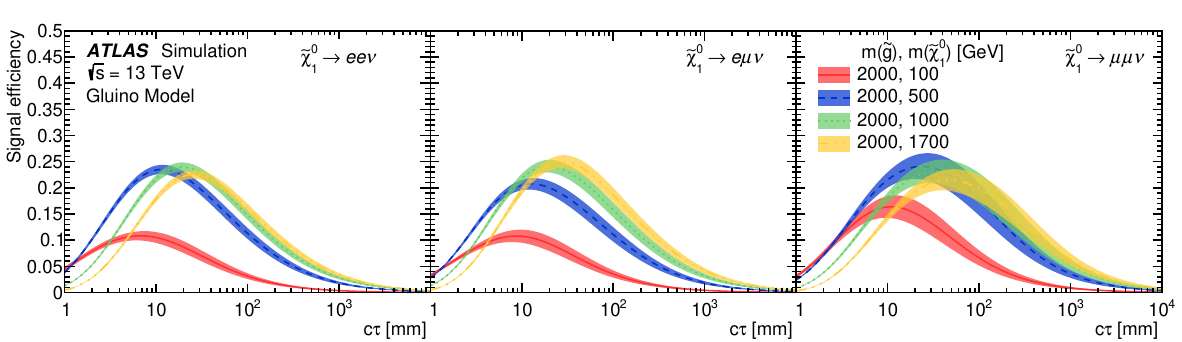}
\caption{Total signal efficiency for an LLP decay as a function of $c\tau$ for the gluino-produced $\tilde{\chi}_{1}^{0}$ with a $\tilde{g}$ mass of 2000 \GeV and $\tilde{\chi}_{1}^{0}$ mass of 100, 500, 1000, or 1700 \GeV decaying into (left) $ee\nu$, (middle) $e\mu\nu$, and (right) $\mu\mu\nu$ final states. The colored bands represent the total uncertainty at one standard deviation.}
\label{fig:tot_vtx_2000_1000_trio}
\end{figure}

Figure~\ref{fig:tot_vtx_2000_1000_trio} shows an example of the efficiency for a $\tilde{\chi}_{1}^{0}$ decay to satisfy the full selection criteria as a function of $c\tau$.
The uncertainty band represents the total uncertainty at one standard deviation, including the sources of systematic uncertainty discussed in Section~\ref{sec:systs}.
The extrapolation to lifetimes other than those of the generated samples is discussed in Section~\ref{sec:result}.
The trigger efficiency for dimuon vertices is lower than that for vertices with electrons, but its impact on the total dimuon efficiency is mitigated by the higher reconstruction efficiency for displaced muons previously shown in Figure~\ref{fig:modifiedID}.
For the same parent mass, the peak efficiency for higher LLP masses occurs at larger proper lifetimes due to the lower boost of the LLP\@.
A highly boosted LLP, such as in the case of the 100 \GeV $\tilde{\chi}_{1}^{0}$ and 2000 \GeV $\tilde{g}$, can produce highly collimated daughters, resulting in the reconstruction of only one lepton and hence lower signal efficiency.
The efficiencies and systematic uncertainties are presented per LLP decay for each vertex type; however, all benchmark samples contain two LLPs and hence limits are evaluated in Section~\ref{sec:result} assuming two decays per event.

It is of particular interest to compare the reconstruction efficiency shown in Figure~\ref{fig:tot_vtx_2000_1000_trio} with that obtained by ATLAS for displaced dimuon vertices reconstructed using standalone muons~\cite{EXOT-2017-03}, i.e. muons reconstructed using MS information only.
While the two analyses probe different RPV SUSY models, the comparison nevertheless provides useful context on the complementarity of ID and MS displaced vertex searches.
When using standalone muons, the efficiency for events containing a pair of long-lived $\tilde{\chi}_{1}^{0}$ (each with $m({\tilde{\chi}_{1}^{0}})=1000$~\gev) peaks at $\sim$10\% at a $c\tau$ of 1000~mm.
The current analysis, which includes ID tracks and uses a more sophisticated DV reconstruction algorithm, reaches a higher peak efficiency at lower $c\tau$.
When considering a same-mass $\tilde{\chi}_{1}^{0}$ and only considering a single decay into a muon pair, this analysis achieves an efficiency that peaks at $\sim$23\% at a $c\tau$ of 34~mm and yields only a somewhat lower value of $\sim$6\% at 1000~mm.

\section{Background estimate}
\label{sec:bkg}

The background from decays of SM particles produced by proton--proton collisions is reduced to negligible levels by requirements on the invariant mass of the signal lepton pair forming the DV and the transverse displacement of the DV relative to the PV\@.
Two additional sources of background are considered: DVs resulting from cosmic-ray muons and from randomly crossing lepton tracks.

The dimuon cosmic-ray background is estimated in a control region of dimuon DVs satisfying all signal requirements except for the cosmic veto, which is inverted.
The tail of the $\rco$ distribution of dimuon DVs in this region ($\rco<0.01$) is fit to an exponential distribution and extrapolated into the signal region ($\rco>0.01$).
The fit range, which starts at $\rco=0.0027$, is chosen to produce the best agreement between the prediction and observed DVs in the region $0.005<\rco<0.01$, adjacent to the signal region.
The extrapolation yields an estimated cosmic background of $(8.3\pm6.0)\times{}10^{-3}$ $\mu^+\mu^-$ events, where the uncertainty is statistical.

One of the two reconstructed muon tracks from a cosmic-ray muon may fail to satisfy the muon signal requirements while satisfying the electron signal requirements due to bremsstrahlung by the muon, so the cosmic veto is also applied to electron-muon pairs.
A similar extrapolation method cannot be used to estimate a cosmic background in the $e^{\pm}\mu^{\mp}$ signal region because there are only two $e^{\pm}\mu^{\mp}$ events with $\rco<0.01$.
The $\rco$ distribution is instead assumed to be similar to that of $\mu^{+}\mu^{-}$ events but suppressed by a factor equal to the ratio of the statistics of the two control regions.
The dimuon cosmic background is thus scaled down to yield an estimated background of $(1.2\pm0.9)\times10^{-5}$ $e^{\pm}\mu^{\mp}$ vertices in the signal region, where the uncertainty is statistical.

The background from two randomly crossing lepton tracks forming a signal DV is estimated from data by using large numbers ($10^6-10^8$, depending on the channel) of toy ``mixed'' events.
Each toy event consists of two randomly selected leptons (which satisfy all signal lepton requirements) from different data events with a randomly selected PV from a third event.
The lepton pair is only considered a ``candidate pair'' if it has the necessary kinematics to be selected by one of the triggers, the leptons have opposite-sign charges, and the invariant mass of the leptons is greater than 12~\GeV.
The lepton tracks are shifted spatially around the new PV to preserve the impact parameters relative to the PV in their original events.
Due to computational limitations, these tracks retain their pattern of hits in each layer even if the shifting would result in a different expected hit pattern.
DV reconstruction is then run on each toy event and, if a DV is successfully reconstructed, the analysis requirements are applied to the DV\@.
This method yields the probability for a signal DV to be produced from two randomly crossing leptonic tracks.
This probability can then be applied to the number of candidate lepton pairs in the data.
Candidate pairs in data include any pairs of ID tracks in the same event that satisfy the same criteria used in toy selection.

Multiple validations of the event-mixing process were performed on signal MC, non-signal MC ($t\bar{t}$), and data.
The spatial shifting of tracks was validated by shifting all tracks in an MC signal event around a randomly selected PV and rerunning the vertex reconstruction algorithm.
The overall difference between the number of reconstructed signal DVs due to shifting tracks is 1.1\%.
There is a loss of ${\sim}3$ \% of signal DVs, but a gain of ${\sim}2$\% of signal DVs that previously were not reconstructed, yielding a net loss of 1.1\%.
A vertex that originally failed to satisfy the selection criteria because it was in a disabled module but was ``shifted’’ out may now satisfy the criteria, while vertices can be lost to the same effect.
The largest effect is due to rejecting individual tracks due to differences between the expected patterns of hits in the silicon layers.
For example, a track that originally intersects with a dead pixel in the next layer outside the DV (thus fulfilling the DV reconstruction algorithm requirements) will lack the expected hit if moved into a region without an adjacent dead pixel, resulting in rejection by the DV reconstruction algorithm.
The predictive ability of the method was tested using non-leptonic tracks (see Section~\ref{sec:systs}), which is used to set a systematic uncertainty.

There are 6646 $e^{+}e^{-}$, 792 $e^{\pm}\mu^{\mp}$, and 3134 $\mu^{+}\mu^{-}$ candidate pairs observed in data, and at least $10^3$ times more toy candidate pairs were generated for each vertex type.
None of the reconstructed dilepton DVs in the toy events satisfied all signal requirements.
The upper limit (at 95\% confidence) of the probability of randomly crossing leptons producing a signal DV is $\mathcal{O}(10^{-6})$ for each of $e^{+}e^{-}$, $e^{\pm}\mu^{\mp}$, and $\mu^{+}\mu^{-}$.
Rather than using a fixed probability, the numbers of toy and observed candidate pairs are treated as control and signal regions respectively when setting limits on signal cross-sections, allowing for the probability to be fit simultaneously with the signal strength.
The procedure for determining the limits is detailed in Section~\ref{sec:result}.


\section{Systematic Uncertainties}
\label{sec:systs}

The dominant source of systematic uncertainty in the signal efficiency is the modeling of track and DV reconstruction.
This uncertainty is estimated from a sample of $K_{S}\rightarrow\pi^+\pi^-$ reconstructed in MC and data.
Due to the relatively low mass and $p_{\mathrm{T}}$ of the $K_{S}$, few of the signal requirements can be applied beyond the standard vertex quality requirements.
Instead, to select $K_{S}$, the DVs are required to have exactly two oppositely charged-particle tracks, 470~\MeV~$<m_{\pi\pi}<$~530~\MeV, $R_{xy}$~$>15$~mm, and tracks with longitudinal separation $|\Delta z_{0}|<2$~mm.
The number of reconstructed $K_{S}$ vertices is binned in $R_{xy}$ after reweighting the simulated events such that the $p_{\mathrm{T}}$ spectrum of the $K_{S}$ in the MC reproduces that in the data.
The vertices are categorized based on the algorithms that reconstructed the two tracks from the $K_{S}$ decay (ST/ST, ST/LRT, and LRT/LRT) with separate systematic uncertainties calculated for each category.
A common normalization for the three categories is used by normalizing the number of ST/ST vertices at low $R_{xy}$ in the MC to the data, where modeling is best understood for both track and vertex reconstruction.
The residual discrepancy between MC and data in each bin is taken as the systematic uncertainty.
Some variation is observed in the agreement between data and MC based on the year and corresponding MC sample considered.
Consequently, a separate normalization factor and systematic uncertainty is calculated for each year.
The systematic uncertainty is generally less than $20\%$ per vertex at low and moderate $R_{xy}$, but at high $R_{xy}$ it can be up to $34\%$ for ST/ST vertices, $40\%$ for ST/LRT vertices, and $44\%$ for LRT/LRT vertices.

Using the residual discrepancy between MC and data as the systematic uncertainty for reconstructing a DV results in a conservative estimate in comparison to decomposing the uncertainty into a correlated and uncorrelated component relative to the reconstruction of the two tracks.
Inefficiencies in $K_{S}$ reconstruction may result from the failure to reconstruct individual tracks due to effects such as multiple scatterings, or from a failure to reconstruct both tracks such as when the particle decays within a dead Si module.
The former is uncorrelated between tracks, while the latter is an example of a fully correlated effect.
The residual discrepancy may be converted into uncorrelated, per-track uncertainties and propagated through to vertices~\cite{SUSY-2018-13,EXOT-2018-57}, which will yield a smaller uncertainty for the vertex.
Possible correlations can then be estimated via expanded studies of the geometric or time dependence of the data/MC ratio.
The statistical analysis discussed in Section~\ref{sec:result} does not have a strong dependence on this dominant uncertainty; therefore the simpler but conservative method that assumes full correlation is used.

\begin{figure}[h]
\centering
\includegraphics[width=1.0\textwidth]{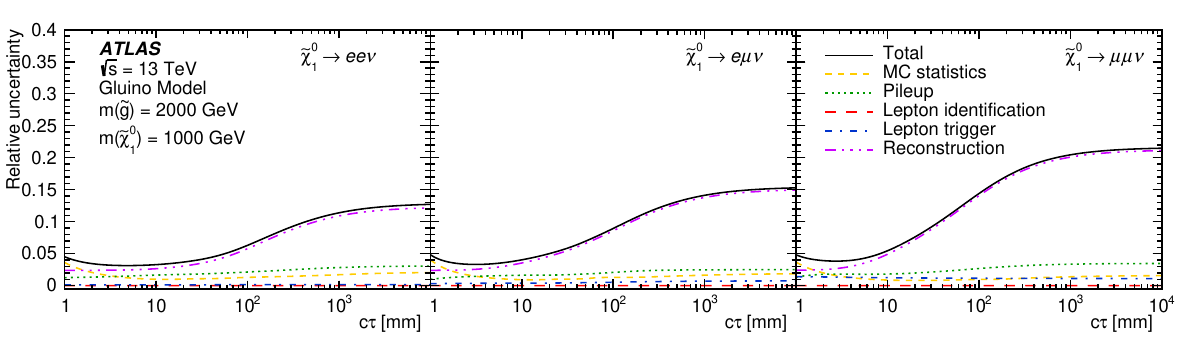}

\caption{Systematic uncertainties in the signal DV selection efficiency extrapolated as a function of $c\tau$ for the gluino-produced $\tilde{\chi}_{1}^{0}$ with a $\tilde{g}$ mass of 2000 \GeV and $\tilde{\chi}_{1}^{0}$ mass of 1000 \GeV decaying into (left) $ee\nu$, (middle) $e\mu\nu$, and (right) $\mu\mu\nu$ final states.}
\label{fig:tot_vtx_2000_err_trio}
\end{figure}

Highly boosted and collimated lepton pairs may have a small $d_0$ even when the $R_{xy}$ of their DV is large, resulting in a significant proportion of displaced ST/ST and ST/LRT vertices for some mass points.
The systematic uncertainty is therefore calculated separately for each mass hypothesis.
Figure~\ref{fig:tot_vtx_2000_err_trio} shows the extrapolated average systematic uncertainties per vertex for a benchmark sample as a function of $c\tau$.
The systematic uncertainty increases at higher $c\tau$ due to the increasing presence of LRT tracks.
The reconstruction efficiency for muons with LRT tracks is higher than that for electrons with LRT tracks as shown in Figure~\ref{fig:modifiedID}.
Consequently, at large $c\tau$, where more LRT tracks are produced, both the average efficiency (Figure~\ref{fig:tot_vtx_2000_1000_trio}) and the average systematic uncertainty (Figure~\ref{fig:tot_vtx_2000_err_trio}) are higher for muons than for electrons.

Custom multiplicative factors are calculated as a function of lepton kinematics to correct MC lepton distributions for the imperfect modeling of the efficiency of the triggers and lepton identification working points.
The factors are calculated using a tag-and-probe method similar to that described in~\cite{EGAM-2021-01} based on $Z\rightarrow \ell^+\ell^-$ events in data and MC\@.
The multiplicative factors for electron triggering and identification are approximately $1$, except in the transition region between the barrel and endcap calorimeters ($1.37 < |\eta| < 1.52$) where values for triggering generally range from 0.90 to 1.02 and values for identification range from 0.98 to 1.02.
For muon identification, values are likewise approximately $1$ except in the central $\eta$ region, where they can be as low as $0.98$.
For triggering on muons, the multiplicative factors typically range from 0.7 to 1.0.
The factors have an associated uncertainty that is propagated through the analysis by shifting the factors upwards and downwards by their associated systematic uncertainty.
The final systematic uncertainties associated with the triggers and lepton identification are both approximately $1\%$ or smaller.
The theory uncertainties associated with modeling the benchmark signal samples are the uncertainty in the cross-section, the PDF uncertainties, the uncertainties in the factorization and renormalization scales, and the uncertainty associated with initial-state and final-state radiation combined with multiple parton interactions (ISR/FSR/MPI).
The methods used follow those described in Ref.~\cite{Borschensky:2014cia}.
The largest of these uncertainties are the PDF uncertainties, at 5.2\%, followed by the ISR/FSR/MPI uncertainty at 3.0\%.
These values are significantly smaller than the dominant sources of systematic uncertainties so the largest value calculated amongst the samples is applied to all signal samples as a constant value.

The systematic uncertainty associated with the random track crossing background is calculated using ID tracks not identified as leptons but satisfying all other signal requirements.
Due to computational limitations, shifted tracks retain their original set of hits in the ID tracking layers, even if the shift would result in a different expected hit pattern.
The systematic impact from this is inherently included in the systematic uncertainty determined from comparing observed and predicted number of non-leptonic vertices.
The DV mass and $R_{xy}$ distributions observed in data are well reproduced by the DVs created from toy events.
The probability of randomly crossing non-leptonic tracks to form a signal DV is calculated to be $(3.07 \pm 0.09) \times 10^{-7}$.
That probability is applied to the observed number of non-leptonic track candidate pairs in data ($4.22 \times 10^{11}$) to yield the predicted number of non-leptonic DVs in data of $143800 \pm{} 4200$.
The observed number of non-leptonic DVs is $126210$.
The difference between the expectation and observation is combined in quadrature with the statistical uncertainty to yield a systematic uncertainty of 14\%.

The systematic uncertainty associated with the procedure of fitting an exponential fit function to the cosmic background histogram is estimated by varying the histogram bin width and varying the lower bound of the fit range.
The two contributions are combined in quadrature to yield a systematic uncertainty of 44\%.

The uncertainty in the combined 2015--2018 integrated luminosity is 0.83\%~\cite{DAPR-2021-01}, obtained using the LUCID-2 detector~\cite{LUCID2} for the primary luminosity measurements, complemented by measurements using the inner detector and calorimeters.
The uncertainty associated with the pileup weighting procedure was evaluated by applying alternate weight sets and found to increase with higher $c\tau$, while remaining sub-dominant.

Finally, the uncertainty associated with the lifetime extrapolation procedure described below was calculated by comparing the deviation of nominal efficiencies to extrapolated values.
A 2.0\% uncertainty is conservatively assigned to the signal DV selection efficiency.


\section{Results}
\label{sec:result}

No events are observed in the signal region, which is consistent with the background expectation.
The analysis is a cut-and-count analysis, allowing limits to be set for each benchmark model using a single bin counting experiment.
Therefore, limits are set at the 95\% confidence level using the CLs method~\cite{Read:2002hq} based on RooStats~\cite{RooFit} via a toys-based method with 25,000 pseudo-experiments.
The signal region includes both the observed number of events (zero), the expected number of signal events for each benchmark model, and the cosmic-ray muon background estimates for $\mu\mu$ and $e\mu$ vertices.
Additionally, the numbers of candidate track pairs discussed in Section~\ref{sec:bkg} is included.
Three control regions, one for each vertex type, are defined to include the number of candidate pair toy events created and the number of signal vertices created in the toy events (zero).
While performing the fit to observed data, both the signal strength and the probability of producing a signal vertex from randomly crossing track pairs can be scaled, allowing for the possibility of non-zero probabilities despite the lack of toy DVs in the signal region.
Due to the lack of observed events after unblinding, the fitted probabilities are generally $O(10^{-10})$, with an error several orders of magnitude larger than the central values.
When accounting for the error, the order of magnitude is comparable to but does not exceed the calculated upper limit on the crossing probability of $O(10^{-6})$ discussed in Section~\ref{sec:bkg}.

Limits are set separately for each decay channel as a function of $c\tau$, utilizing a lifetime reweighting procedure applied to signal efficiencies.
The lifetime reweighting procedure extrapolates the acceptance times efficiency of the finite number of signal samples generated with specific $c\tau$ to a continuous distribution.
Weights are applied to each LLP decay vertex corresponding to the relative frequency of decays at the lifetime of that individual particle as a function of the generated and target sample $c\tau$.
More details can be found in Ref.~\cite{SUSY-2017-04}.

Only one LLP decay must be detected for the event to qualify as a signal event.
The production of two LLPs per event in the benchmark models studied increases the probability that at least one signal vertex is found.
For the $Z'$ benchmark samples the limits are presented at event-level, assuming both $Z'$ decay into the same vertex type ($e^{+}e^{-}$, $e^{\pm}\mu^{\mp}$, or $\mu^{+}\mu^{-}$).
For the SUSY models the limits are interpreted assuming a coupling of $\lambda_{121} \neq 0$ or $\lambda_{122} \neq 0$ (with all other $\lambda_{ijk} = 0$), resulting in decays into $ee$ and $e\mu$ in the former case and $e\mu$ and $\mu\mu$ in the latter.
Extracted limits are shown together with the theory cross-section for the SUSY models.
Theory signal cross-sections for the Gluino model are calculated to approximate next-to-next-to-leading order in the strong coupling constant, adding the resummation of soft gluon emission at next-to-next-to-leading-logarithmic accuracy (approximate NNLO+NNLL)~\cite{Beenakker:2016lwe,Beenakker:2014sma,Beenakker:2013mva,Beenakker:2011sf,Beenakker:2009ha,Kulesza:2009kq,Kulesza:2008jb,Beenakker:1996ch}. The nominal cross-section and the uncertainty are derived using the PDF4LHC15\_mc PDF set, following the recommendations of Ref.~\cite{Butterworth:2015oua}.
Theory signal cross-sections for the EWkino model are calculated to next-to-leading order in the strong coupling constant, adding the resummation of soft gluon emission at next-to-leading-logarithmic accuracy (NLO+NLL)~\cite{Beenakker:1999xh,Bozzi:2007qr,Fuks:2013vua,Fuks:2013lya,Fiaschi:2018xdm}. The nominal cross-section and the uncertainty are taken from an envelope of cross-section predictions using different PDF sets and factorization and renormalization scales, as described in Ref.~\cite{Borschensky:2014cia}.

\begin{figure}[tbp]
\centering
\includegraphics[width=1.0\textwidth]{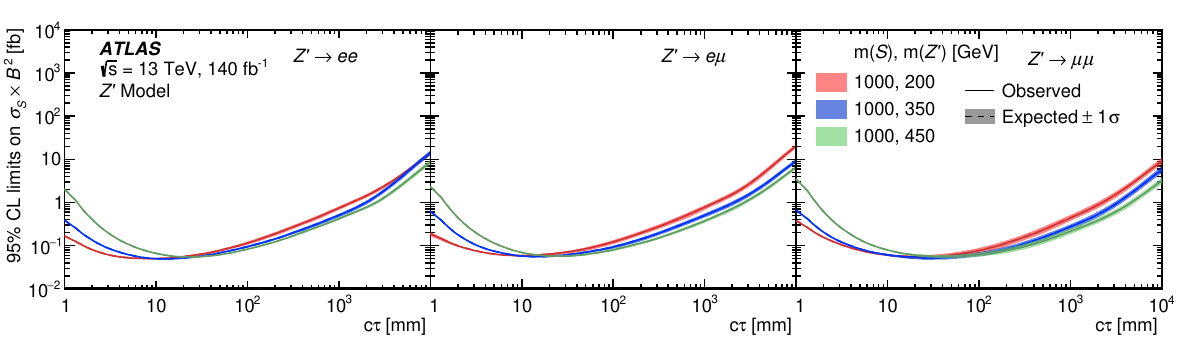}

\caption{Observed and expected limits on the product of cross-section and ${B}^{2}$ as a function of $c\tau$ for models with an $S$ mass of 1000~\GeV and $Z'$ masses between 200~and~450~\GeV decaying into (left) $e^+e^-$, (middle) $e^{\pm}\mu^{\mp}$, and (right) $\mu^{+}\mu^{-}$ final states. The branching ratio $B$ is assumed to be 100\% to the given final state for both $Z'$.}
\label{fig:lim_zprime_1000}
\end{figure}

Observed and expected limits on the $Z'$ production cross-section times branching ratio as a function of lifetime are shown in Figure~\ref{fig:lim_zprime_1000} for  $m_\mathrm{S}=1000~\GeV$.
Both the observed and expected limits are plotted, but the lines overlap due to the agreement between the background prediction and observation.
At longer $Z'$ lifetimes, weaker limits are set on lighter $Z'$ for the same scalar mass because the more highly boosted $Z'$ results in more decays outside of the ID\@.
Conversely, at shorter lifetimes the limits are weaker for heavier $Z'$ because the less-boosted particles will often decay before reaching the ID\@.
The upper limits for different scalar masses are similar with stronger limits set for higher scalar masses due to the higher reconstruction efficiency.
For example, the upper limits for the $e\mu$ channel at 10~mm (near the peak efficiencies) are 0.1120,~0.0589,~and~0.0534~fb for ($m_S$,~$m_{Z'}$)~=~(500,~200)~\GeV, (1000,~200)~\GeV, and (1500,~250)~\GeV respectively.
The larger error bands for dimuon vertices are a result of the larger track and DV reconstruction systematic uncertainty for these vertices, as discussed in Section~\ref{sec:systs}.

\begin{figure}[tbp]
\centering
\includegraphics[width=1.0\textwidth]{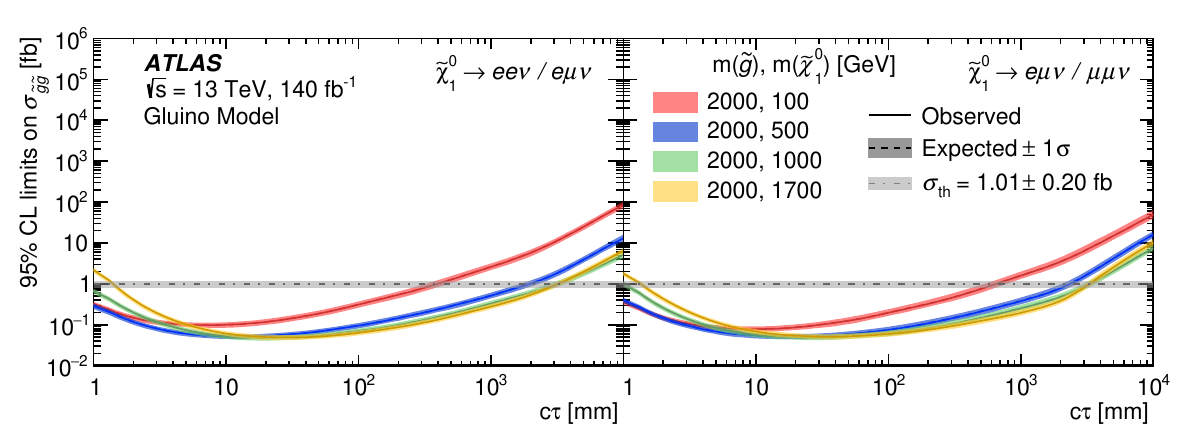}

\caption{Observed and expected limits on the signal production cross-section as a function of $c\tau$ for the gluino-produced $\tilde{\chi}_{1}^{0}$ with a $\tilde{g}$ mass of 2000~\GeV and $\tilde{\chi}_{1}^{0}$ masses of 100,~500,~1000,~and~1700~\GeV with the $\tilde{\chi}_{1}^{0}$ decaying with equal branching ratios to (left) $ee\nu$ and $e\mu\nu$ and to (right) $e\mu\nu$ and $\mu\mu\nu$. These scenarios correspond to RPV couplings (left) $\lambda_{121} \neq 0$ or (right) $\lambda_{122} \neq 0$ with all other $\lambda_{ijk}$ assumed to be zero. The dashed horizontal line represents the theory production cross-section ($\sigma_{\mathrm{th}}$).}
\label{fig:lim_susy_gg_2000}
\end{figure}

Observed and expected limits on the gluino-produced $\tilde{\chi}_{1}^{0}$ signal cross-section as a function of $c\tau$ are shown in Figure~\ref{fig:lim_susy_gg_2000} for $m(\tilde{g}) = 2000$~\GeV.
As in the case of the $Z'$ model limits, the lines for the observed and expected limits overlap in both SUSY models.
For the same $\tilde{g}$ mass, more stringent limits are set at long lifetimes for heavier $\tilde{\chi}_{1}^{0}$ masses and at shorter lifetimes for lighter $\tilde{\chi}_{1}^{0}$ masses---the same parent-daughter trend observed for the $Z'$ samples.
For $m(\tilde{g}) = 2000$~\GeV, the largest excluded $c\tau$ in the $\lambda_{121}$ ($\lambda_{122}$) channel ranges from 400~mm (660~mm) for $m(\tilde{\chi}_{1}^{0}) = 100$~\GeV up to 3200~mm (3300~mm) for $m(\tilde{\chi}_{1}^{0}) = 1000$~\GeV.
The smallest lifetime excluded is the 1~mm lower bound of the range investigated in this analysis for all $\tilde{\chi}_{1}^{0}$ masses except for $m(\tilde{\chi}_{1}^{0}) = 1700$~\GeV, which is only excluded down to $\sim$1.5~mm in both channels.
For the higher mass $m(\tilde{g}) = 2500$~\GeV, a narrower range of lifetimes is excluded, with no exclusion for $m(\tilde{\chi}_{1}^{0}) = 100$~\GeV.
For $\lambda_{121}$, the excluded $c\tau$ range for a  $\tilde{\chi}_{1}^{0}$ with a mass of 500,~1000,~or~2200~\gev is between 2.8~and~63~mm, 4.0~and~100~mm, and 7.8~and~150~mm, respectively.
In the case of $\lambda_{122}$, the excluded $c\tau$ range for a  $\tilde{\chi}_{1}^{0}$ with a mass of 500,~1000,~or~2200~\gev is between 3.7~and~92~mm, 4.9~and~150~mm, and 9.1~and~200~mm, respectively.
By contrast, for the lowest gluino mass ($m(\tilde{g}) = 1500$~\GeV) and in both channels, all $c\tau$ between 1~and~10\,000~mm are excluded for all $\tilde{\chi}_{1}^{0}$ masses except for 100~\gev, where the largest excluded lifetime is at 3900~mm (5400~mm) for the $\lambda_{121}$ ($\lambda_{122}$) channel.

\begin{figure}[tbp]
\centering
\includegraphics[width=1.0\textwidth]{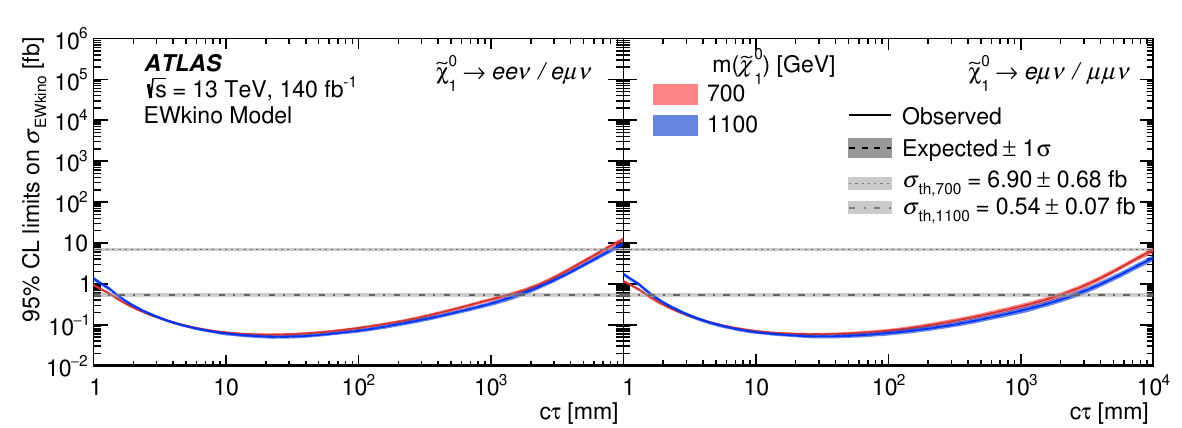} \

\caption{Observed and expected limits on the signal production cross-section as a function of $c\tau$ for the electroweakino-produced $\tilde{\chi}_{1}^{0}$ with masses of 700~and~1100~\GeV with the $\tilde{\chi}_{1}^{0}$ decaying with equal branching ratios to (left) $ee\nu$ and $e\mu\nu$ and to (right) $e\mu\nu$ and $\mu\mu\nu$. These scenarios correspond to RPV couplings (left) $\lambda_{121} \neq 0$ or (right) $\lambda_{122} \neq 0$ with all other $\lambda_{ijk}$ assumed to be zero. The dashed horizontal lines represent the theory production cross-sections ($\sigma_{\mathrm{th}}$).}
\label{fig:lim_susy_ew_1000}
\end{figure}

Observed and expected limits on an electroweakino-produced $\tilde{\chi}_{1}^{0}$ signal cross-section as a function of lifetime are shown in Figure~\ref{fig:lim_susy_ew_1000} for $m(\tilde{\chi}_{1}^{0}) = 700$~\GeV and $m(\tilde{\chi}_{1}^{0}) = 1100$~\GeV.
For $m(\tilde{\chi}_{1}^{0}) = 1100$~\GeV, $c\tau$ are excluded with a lower bound of 1.5~mm (1.7~mm) and upper bound of 1600~mm (2600~mm) for the $\lambda_{121}$ ($\lambda_{122}$) channel.
Limits are more modest for $m(\tilde{\chi}_{1}^{0}) = 1300$~\GeV, ranging between about $3.1$~mm ($3.6$~mm) for the lower bound and 460~mm (660~mm) for the upper bound for the $\lambda_{121} \neq 0$ ($\lambda_{122} \neq 0$) channel.
For $m(\tilde{\chi}_{1}^{0}) = 900$~\GeV and $m(\tilde{\chi}_{1}^{0}) = 700$~\GeV the lower bound of the limit on $c\tau$ is at 1~mm, the lower bound of the investigated range.
In the case of $m(\tilde{\chi}_{1}^{0}) = 900$~\GeV the excluded $c\tau$ range extends to 3300~mm (5200~mm) for the $\lambda_{121}$ ($\lambda_{122}$) channel, while for $m(\tilde{\chi}_{1}^{0}) = 700$~\GeV the range extends to a larger 7200~mm for $\lambda_{121}$ and to the edge of the evaluated range for $\lambda_{122}$.
For $m(\tilde{\chi}_{1}^{0}) = 500$~\GeV and lighter, the full range of considered $c\tau$ is excluded.

%
%
%

%
\section{Conclusion}
\label{sec:conclusion}

A search for long-lived neutral particles decaying into pairs of leptons was performed using data collected with the ATLAS detector at the LHC\@.
This is the first analysis to search for heavy LLP decays into displaced dilepton vertices using the full Run-2 $\mathit{pp}$ data.
Previous searches by the ATLAS and CMS Collaborations focused on lighter LLPs (usually decaying into dimuon pairs) or used partial Run-2 data.
No previous analysis, other than the ATLAS search using similar reconstruction and selection criteria~\cite{SUSY-2017-04} but less than one third of the data, has searched for all three signatures ($ee$, $e\mu$, and $\mu\mu$ DVs).

No events were observed in the signal region, which is consistent with background expectations.
These results complement the previous search based on 32.8~\ensuremath{\text{fb}^{-1}} of $\mathit{pp}$ data\xspace~\cite{SUSY-2017-04} interpreted using different models.
The search yields new limits on the cross-section for the production of a scalar that decays into a pair of long-lived $Z'$, both decaying into the same final state ($e^{+}e^{-}$, $e^{\pm}\mu^{\mp}$, or $\mu^{+}\mu^{-}$), with $c\tau$ from 1~to~10\,000~mm.
The extracted limits are for a scalar with mass of 0.5~to~1.5~\TeV and $Z'$ with mass of 100~to~700~\GeV.
The search also excludes the production of a long-lived neutralino $\tilde{\chi}_{1}^{0}$ that decays through a $\lambda_{121}$ or $\lambda_{122}$ coupling into two leptons and a neutrino in a similar lifetime range.
From the pair production of gluinos with mass of 1.5~to~2.0~\TeV, $\tilde{\chi}_{1}^{0}$ with mass of 0.1~to~1.7~\TeV are excluded at most $c\tau$ from  1~to~10\,000~mm.
For $\tilde{\chi}_{1}^{0}$ produced directly or in the decay of $\tilde{\chi}_{2}^{0}$ or $\tilde{\chi}_{1}^{\pm}$, neutralinos with mass of 0.1~to~0.5~\TeV are excluded for all lifetimes in the considered range, with smaller lifetime ranges excluded for $\tilde{\chi}_{1}^{0}$ masses increasing from 0.7~to~1.3~\TeV.
These are the first limits set on high mass, long-lived neutralino models using gluino and electroweakino production modes with decays into lepton pairs, with the exception of gluino-produced neutralino decays into muons~\cite{EXOT-2017-03}.

\section*{Acknowledgments}
%

%
%

%
%

We thank CERN for the very successful operation of the LHC and its injectors, as well as the support staff at
CERN and at our institutions worldwide without whom ATLAS could not be operated efficiently.

The crucial computing support from all WLCG partners is acknowledged gratefully, in particular from CERN, the ATLAS Tier-1 facilities at TRIUMF/SFU (Canada), NDGF (Denmark, Norway, Sweden), CC-IN2P3 (France), KIT/GridKA (Germany), INFN-CNAF (Italy), NL-T1 (Netherlands), PIC (Spain), RAL (UK) and BNL (USA), the Tier-2 facilities worldwide and large non-WLCG resource providers. Major contributors of computing resources are listed in Ref.~\cite{ATL-SOFT-PUB-2026-001}.

We gratefully acknowledge the support of ANPCyT, Argentina; YerPhI, Armenia; ARC, Australia; BMWFW and FWF, Austria; ANAS, Azerbaijan; CNPq and FAPESP, Brazil; NSERC, NRC and CFI, Canada; CERN; ANID, Chile; CAS, MOST and NSFC, China; Minciencias, Colombia; MEYS CR, Czech Republic; DNRF and DNSRC, Denmark; IN2P3-CNRS and CEA-DRF/IRFU, France; SRNSFG, Georgia; BMFTR, HGF and MPG, Germany; GSRI, Greece; RGC and Hong Kong SAR, China; ICHEP and Academy of Sciences and Humanities, Israel; INFN, Italy; MEXT and JSPS, Japan; CNRST, Morocco; NWO, Netherlands; RCN, Norway; MNiSW, Poland; FCT, Portugal; MNE/IFA, Romania; MSTDI, Serbia; MSSR, Slovakia; ARIS and MVZI, Slovenia; DSI/NRF, South Africa; MICIU/AEI, Spain; SRC and Wallenberg Foundation, Sweden; SERI, SNSF and Cantons of Bern and Geneva, Switzerland; NSTC, Taipei; TENMAK, T\"urkiye; STFC/UKRI, United Kingdom; DOE and NSF, United States of America.

Individual groups and members have received support from BCKDF, CANARIE, CRC and DRAC, Canada; CERN-CZ, FORTE and PRIMUS, Czech Republic; COST, ERC, ERDF, Horizon 2020, ICSC-NextGenerationEU and Marie Sk{\l}odowska-Curie Actions, European Union; Investissements d'Avenir Labex, Investissements d'Avenir Idex and ANR, France; DFG and AvH Foundation, Germany; Herakleitos, Thales and Aristeia programmes co-financed by EU-ESF and the Greek NSRF, Greece; BSF-NSF and MINERVA, Israel; NCN and NAWA, Poland; La Caixa Banking Foundation, CERCA and AGAUR programs from Generalitat de Catalunya and PROMETEO and GenT Programmes Generalitat Valenciana, Spain; G\"{o}ran Gustafssons Stiftelse, Sweden; The Royal Society and Leverhulme Trust, United Kingdom; Eric and Wendy Schmidt Fund for Strategic Innovation, United States of America.

In addition, individual members wish to acknowledge support from Armenia: Yerevan Physics Institute (FAPERJ); CERN: European Organization for Nuclear Research (CERN DOCT); Chile: Agencia Nacional de Investigaci\'on y Desarrollo (ANID FONDECYT reg. 1230987, FONDECYT 1230812, FONDECYT 1240864, Fondecyt 3240661, Fondecyt Regular 1240721); China: Chinese Ministry of Science and Technology (MOST-2023YFA1605700, MOST-2023YFA1609300), National Natural Science Foundation of China (NSFC - 12175119, NSFC 12275265, NSFC-12075060); Czech Republic: Czech Science Foundation (GACR - 24-11373S), Ministry of Education Youth and Sports (ERC-CZ-LL2327, FORTE CZ.02.01.01/00/22\_008/0004632), PRIMUS Research Programme (PRIMUS/21/SCI/017); EU: H2020 European Research Council (ERC - 101002463); European Union: European Research Council (BARD No. 101116429, ERC - 948254, ERC 101089007), Horizon 2020 Framework Programme (MUCCA - CHIST-ERA-19-XAI-00), European Union, Future Artificial Intelligence Research (FAIR-NextGenerationEU PE00000013), Italian Center for High Performance Computing, Big Data and Quantum Computing (ICSC, NextGenerationEU), Marie Sklodowska-Curie Actions (ERC Horizon-2020, GA 956086, GAP-101168829); France: Agence Nationale de la Recherche (ANR-20-CE31-0013, ANR-21-CE31-0013, ANR-21-CE31-0022, ANR-22-EDIR-0002, ANR-24-CE31-0504-01); Germany: Baden-Württemberg Stiftung (BW Stiftung-Postdoc Eliteprogramme), Deutsche Forschungsgemeinschaft (DFG - 469666862, DFG - CR 312/5-2); Italy: Istituto Nazionale di Fisica Nucleare (ICSC, NextGenerationEU), Ministero dell'Università e della Ricerca (NextGenEU 153D23001490006 M4C2.1.1, NextGenEU I53D23000820006 M4C2.1.1, NextGenEU I53D23001490006 M4C2.1.1); Japan: Japan Society for the Promotion of Science (JSPS KAKENHI JP22H01227, JSPS KAKENHI JP22H04944, JSPS KAKENHI JP22KK0227, JSPS KAKENHI JP24K23939); Norway: Research Council of Norway (RCN-314472); Poland: Ministry of Science and Higher Education (IDUB AGH, POB8, D4 no 9722), Polish National Science Centre (NCN 2021/42/E/ST2/00350, NCN OPUS 2023/51/B/ST2/02507, NCN OPUS nr 2022/47/B/ST2/03059, NCN UMO-2019/34/E/ST2/00393, UMO-2020/37/B/ST2/01043, UMO-2022/47/O/ST2/00148, UMO-2023/49/B/ST2/04085, UMO-2023/51/B/ST2/00920); Portugal: Foundation for Science and Technology (FCT); Spain: Generalitat Valenciana (Artemisa, FEDER, IDIFEDER/2018/048, ASFAE/2022/008), Ministry of Science and Innovation (MCIN \& NextGenEU PCI2022-135018-2, MICIN \& FEDER PID2021-125273NB, RYC2019-028510-I, RYC2020-030254-I, RYC2021-031273-I, RYC2022-038164-I), Ministerio de Ciencia, Innovación y Universidades/Agencia Estatal de Investigaci\'on (EU NextGenerationEU (PRTR-C17.I1), PID2022-142604OB-C22); Sweden: Carl Trygger Foundation (Carl Trygger Foundation CTS 22:2312), Swedish Research Council (Swedish Research Council 2023-04654, VR 2018-00482, VR 2021-03651, VR 2022-03845, VR 2022-04683, VR 2023-03403), Knut and Alice Wallenberg Foundation (KAW 2018.0458, KAW 2019.0447, KAW 2022.0358); Switzerland: Swiss National Science Foundation (SNSF - PCEFP2\_194658); United Kingdom: Leverhulme Trust (Leverhulme Trust RPG-2020-004), Royal Society (NIF-R1-231091); United States of America: U.S. Department of Energy (ECA DE-AC02-76SF00515), Neubauer Family Foundation.

%
%

%
%
%
%
%
%

%
%

%
%
%
\printbibliography
\clearpage
 
\begin{flushleft}
\hypersetup{urlcolor=black}
{\Large The ATLAS Collaboration}

\bigskip

\AtlasOrcid[0000-0002-6665-4934]{G.~Aad}$^\textrm{\scriptsize 104}$,
\AtlasOrcid[0000-0001-7616-1554]{E.~Aakvaag}$^\textrm{\scriptsize 17}$,
\AtlasOrcid[0000-0002-5888-2734]{B.~Abbott}$^\textrm{\scriptsize 122}$,
\AtlasOrcid[0000-0002-0287-5869]{S.~Abdelhameed}$^\textrm{\scriptsize 85b}$,
\AtlasOrcid[0000-0002-1002-1652]{K.~Abeling}$^\textrm{\scriptsize 55}$,
\AtlasOrcid[0000-0001-5763-2760]{N.J.~Abicht}$^\textrm{\scriptsize 49}$,
\AtlasOrcid[0000-0002-8496-9294]{S.H.~Abidi}$^\textrm{\scriptsize 30}$,
\AtlasOrcid[0009-0003-6578-220X]{M.~Aboelela}$^\textrm{\scriptsize 45}$,
\AtlasOrcid[0000-0002-9987-2292]{A.~Aboulhorma}$^\textrm{\scriptsize 36e}$,
\AtlasOrcid[0000-0001-5329-6640]{H.~Abramowicz}$^\textrm{\scriptsize 156}$
% ...
\bigskip
\\

% ...
$^{17}$Department for Physics and Technology, University of Bergen, Bergen; Norway.\\
% ...
$^{30}$Physics Department, Brookhaven National Laboratory, Upton NY; United States of America.\\
% ...
$^{36}$$^{(a)}$Facult\'e des Sciences Ain Chock, Universit\'e Hassan II de Casablanca;$^{(b)}$Facult\'{e} des Sciences, Universit\'{e} Ibn-Tofail, K\'{e}nitra;$^{(c)}$Facult\'e des Sciences Semlalia, Universit\'e Cadi Ayyad, LPHEA-Marrakech;$^{(d)}$LPMR, Facult\'e des Sciences, Universit\'e Mohamed Premier, Oujda;$^{(e)}$Facult\'e des sciences, Universit\'e Mohammed V, Rabat;$^{(f)}$Institute of Applied Physics, Mohammed VI Polytechnic University, Ben Guerir; Morocco.\\
% ...
$^{45}$Physics Department, Southern Methodist University, Dallas TX; United States of America.\\
% ...
$^{49}$Fakult\"{a}t Physik, Technische Universit{\"a}t Dortmund, Dortmund; Germany.\\
% ...
$^{55}$II. Physikalisches Institut, Georg-August-Universit\"{a}t G\"ottingen, G\"ottingen; Germany.\\
% ...
$^{85}$$^{(a)}$Khalifa University of Science and Technology, Abu Dhabi;$^{(b)}$New York University Abu Dhabi, Abu Dhabi;$^{(c)}$United Arab Emirates University, Al Ain;$^{(d)}$University of Sharjah, Sharjah; United Arab Emirates.\\
% ...
$^{104}$CPPM, Aix-Marseille Universit\'e, CNRS/IN2P3, Marseille; France.\\
% ...
$^{122}$Homer L. Dodge Department of Physics and Astronomy, University of Oklahoma, Norman OK; United States of America.\\
% ...
$^{156}$Raymond and Beverly Sackler School of Physics and Astronomy, Tel Aviv University, Tel Aviv; Israel.\\
% ...
$^{a}$ Also at Affiliated with an institute formerly covered by a cooperation agreement with CERN.\\
$^{b}$ Also at An-Najah National University, Nablus; Palestine.\\
$^{c}$ Also at Borough of Manhattan Community College, City University of New York, New York NY; United States of America.\\
$^{d}$ Also at Center for High Energy Physics, Peking University; China.\\
$^{e}$ Also at Center for Interdisciplinary Research and Innovation (CIRI-AUTH), Thessaloniki; Greece.\\
$^{f}$ Also at Centre of Physics of the Universities of Minho and Porto (CF-UM-UP); Portugal.\\
% ...

\end{flushleft}


%

%
%

%

%
%
%
%

\end{document}